\documentclass[notitlepage,longbibliography]{revtex4-1}

\usepackage{graphicx}
\usepackage{amsmath}
\usepackage[margin=5pt]{subfig}
\usepackage[usenames]{color}
\definecolor{darkgreen}{rgb}{0.00,0.50,0.25}
\definecolor{darkblue}{rgb}{0.00,0.00,0.67}
\newcommand{\figref}[1]{Fig.~\ref{#1}}
\usepackage[breaklinks,pdftitle={ZeroDB white paper}, pdfauthor={Michael Egorov},colorlinks,urlcolor=blue,citecolor=darkgreen,linkcolor=darkblue]{hyperref}
\usepackage[usenames]{color}
\graphicspath{{pdf/}}

\begin{document}

\title{ZeroDB white paper\footnote{This white paper describes planned functionality but not necessarily the current state of ZeroDB}}

\author{M. Egorov}
\email{michael@zerodb.io}
\author{M. Wilkison}
\email{maclane@zerodb.io}
\affiliation{ZeroDB, Inc.}

\begin{abstract}
ZeroDB is an end-to-end encrypted database that enables clients to operate on (search, sort, query, and share) encrypted data without exposing encryption keys or cleartext data to the database server.
The familiar client-server architecture is unchanged, but query logic and encryption keys are pushed client-side.
Since the server has no insight into the nature of the data, the risk of data being exposed via a server-side data breach is eliminated.
Even if the server is successfully infiltrated, adversaries would not have access to the cleartext data and
cannot derive anything useful out of disk or RAM snapshots.

ZeroDB provides end-to-end encryption while maintaining much of the functionality expected of a modern database, such as full-text search, sort, and range queries.
Additionally, ZeroDB uses proxy re-encryption and/or delta key technology to enable secure, granular sharing of encrypted data without exposing keys to the server and without sharing the same encryption key between users of the database.
\end{abstract}

\date{\today}
\maketitle

\section{Introduction}

Given the cost, performance, and scalability benefits of cloud environments, outsourcing of on-premise infrastructure is accelerating.
However, concerns over data security and the frequency of high-profile data breaches continue to hinder cloud adoption in highly-regulated and security-sensitive industries.
While strong encryption can alleviate many of these concerns by guaranteeing data confidentiality, once data is encrypted it is no longer usable.
Several strategies for computing on encrypted data have recently emerged that seek to provide the benefits of encryption while preserving the functionality of encrypted data.

These techniques fall into two broad categories: (1) encryption schemes that enable server-side computation directly over ciphertext and (2) trusted modules that are assumed to be inaccessible by adversaries~\cite{tutorial}. These two approaches can be further deconstructed into (1) searchable symmetric encryption (SSE), secure multi-party computation (MPC), and fully and partially homomorphic encryption (FHE and PHE, respectively) and (2) client-end solutions and in-cloud solutions, resulting in the taxonomy shown in~\figref{fig:taxonomy}.

\begin{figure}
	\begin{center}
    \subfloat{\label{fig:taxonomy}\includegraphics[width=0.7\columnwidth]{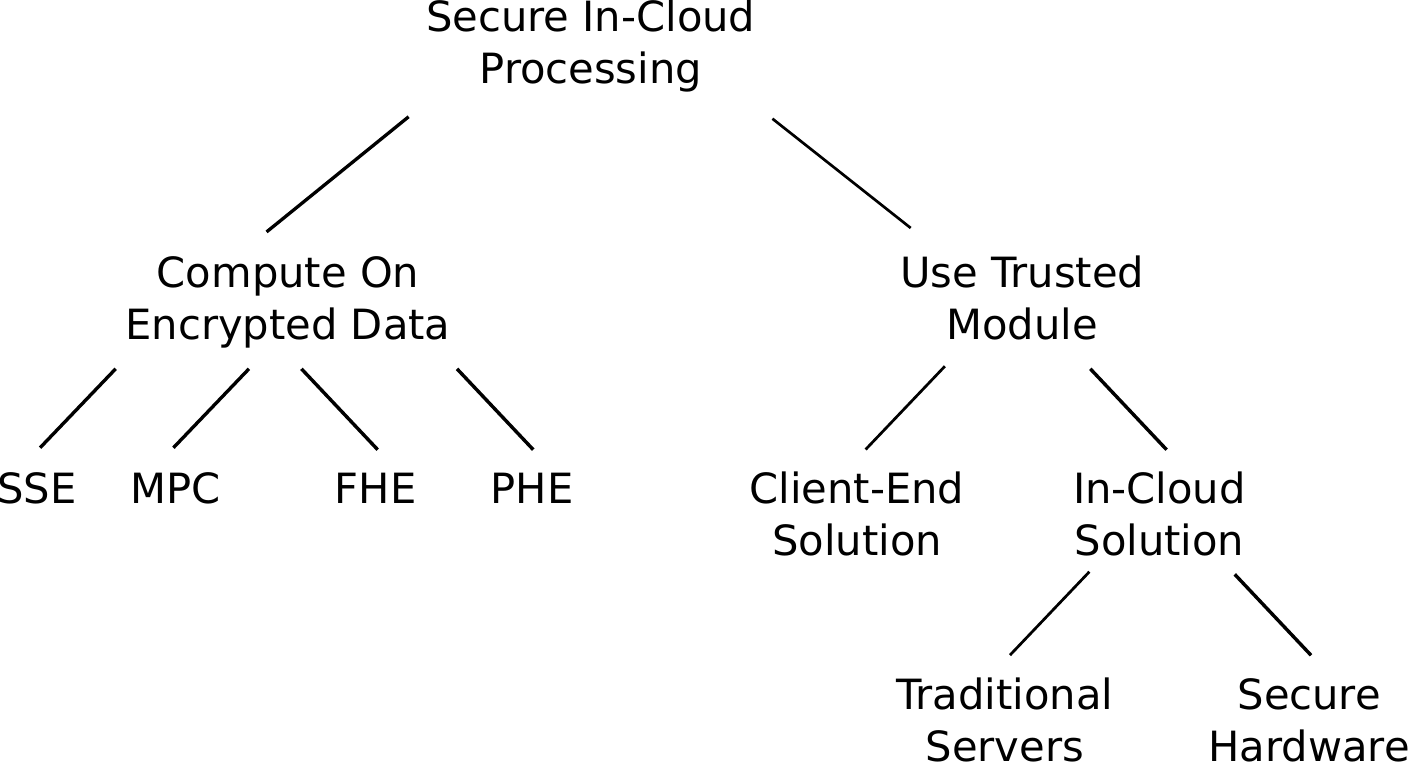}}
	\end{center}
  \caption{Taxonomy of approaches to secure in-cloud processing.
    Computing directly on encrypted data can be done in a fully or partially homomorphic (FHE and PHE, respectively) manner.
    Searchable symmetric encryption (SSE) or secure multi-party computation schemes (MPC) may be used as well.
    Trusted Modules (TM) include client-end and in-cloud solutions.}
	\label{fig:taxonomy}
\end{figure}

Notable works in this area include Gentry's seminal paper on homomorphic encrytion~\cite{gentry}, Popa's CryptDB~\cite{cryptdb}, which executes SQL queries
over encrypted data using a collection of efficient SQL-aware encryption schemes, Dynamic Searchable Symmetric Encryption~\cite{dynamicsse}, Efficient Secure Two-Party Computation Using Symmetric Cut-and-Choose~\cite{mpc}, which yields a 3x performance gain against existing garbled circuit schemes, and Cipherbase~\cite{cipherbase}, a comprehensive database system that provides strong end-to-end data confidentiality through encryption and secure hardware.
Smart provides a more thorough exploration of the latest research and private sector efforts to achieve data encrypted in-use~\cite{smart}.

ZeroDB provides confidentiality of data stored on the database server.
It assumes a client(s) that initially owns the data and associated encryption keys.
This client(s) wishes to outsource the encrypted data to an untrusted server, while maintaining usability of the data.
It assumes passive (honest but curious) adversaries who do not return wrong results to the client.
In the future, we may consider active adversaries and modify our protocol accordingly.

\section{Query protocol}
\label{sec:query-protocol}

The basis of ZeroDB's end-to-end encrypted query protocol is as follows.
The client interacts with the server during the execution of a query over a series of round trips.
An encrypted index is stored on the server as a B-Tree, and the client traverses this index remotely to retrieve the necessary encrypted records.
This remote traversal happens incrementally, with each round trip corresponding to a step down the B-Tree index.
The index consists of buckets which are encrypted before being uploaded to the server and only decrypted client-side.

\subsection{Equality query (using an example of keyword search)}
\begin{figure}
	\begin{center}
        \subfloat[Encrypted index traversal example (simple keyword search)]{\label{fig:tree-traversal}\includegraphics[width=0.47\columnwidth]{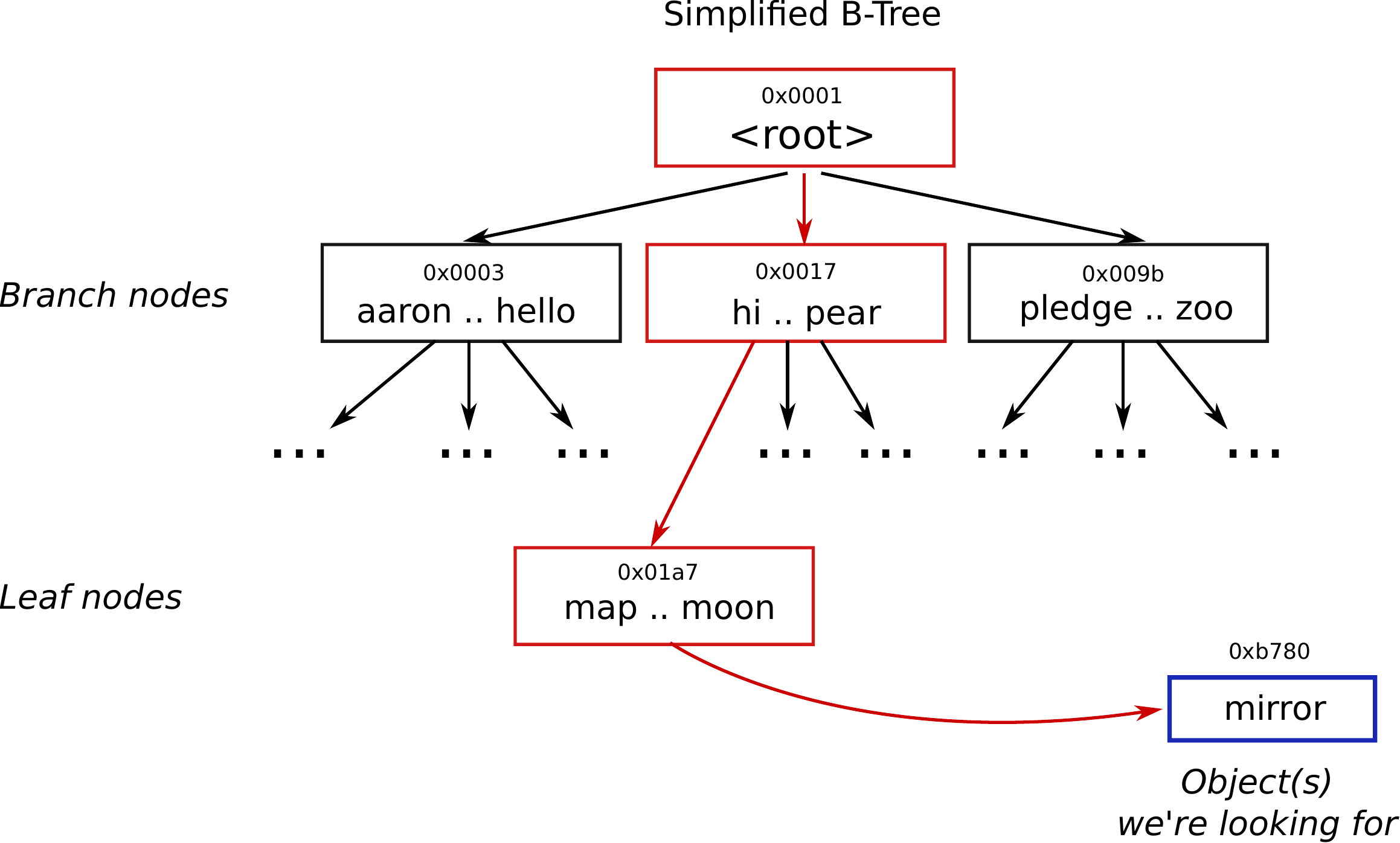}}
        \qquad
        \subfloat[Sequence of client requests for traversal of the example index]{\label{fig:communication}\includegraphics[width=0.3\columnwidth]{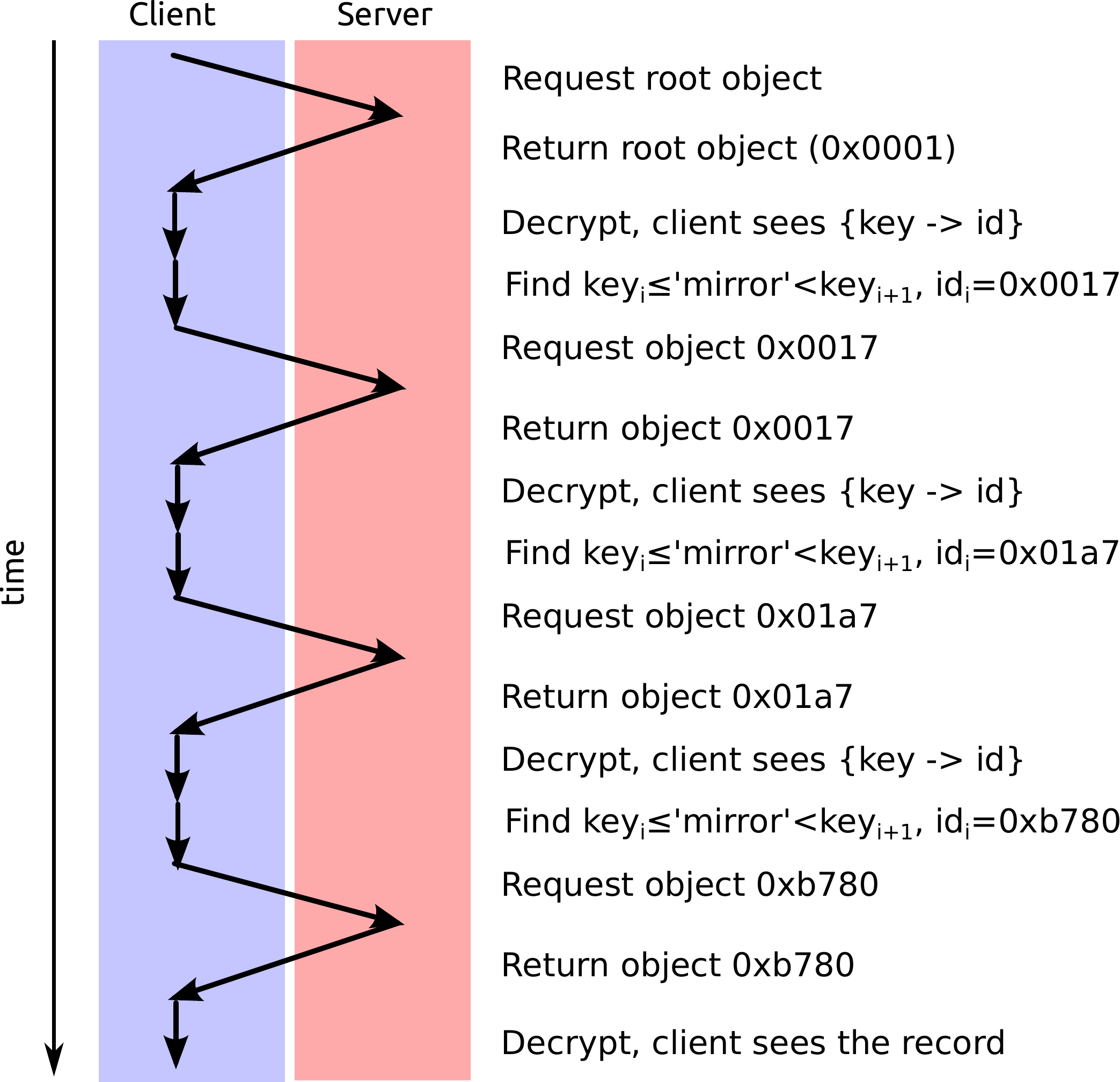}}
	\end{center}
    \caption{Search protocol for equality query using an example of keyword search}
	\label{fig:btree-protocol}
\end{figure}

In ZeroDB, indexes are structured as encrypted B-Trees as in~\figref{fig:tree-traversal}.
A B-Tree consists of encrypted buckets, each of which can be either a root, branch, or leaf node.
The leaf nodes of a tree point to the actual objects being stored.
Thus, searching the database is a simple tree traversal.

In order to make the database end-to-end encrypted yet still capable of performing queries, the client encrypts the buckets (at the time of creation or modification).
The server, which stores the buckets, never knows the encryption key used.
The objects referenced by the leaf nodes of the B-Tree indexes are also encrypted client-side.
As a result, the server doesn't know how individual objects are organized within the B-Tree or whether they belong to an index at all.
Since ZeroDB is encryption agnostic, probabilistic encryption can ensure that the server cannot even compare objects for equality.

When a client performs a query, it asks the server to return buckets of the tree as it traverses the index remotely and incrementally, as in~\figref{fig:communication}.
The client fetches and decrypts buckets in order to figure out which buckets in the next level of the tree to fetch next.
Frequently accessed buckets can be cached client-side so that subsequent queries do not make unnecessary network calls.

The server can infer which level of the tree buckets belong to by observing search patterns,
however it cannot figure out the ordering of elements since it can't see which link from the bucket was used.

The server is responsible for data replication, multi-version concurrency, object locking, user authentication, quotas, and optional immutability/write-once-read-many (WORM) guarantees.
The client performs encryption/decryption and executes query logic.
Thus, even if the server maliciously gives data belonging to user Alice to user Eve,
Eve will not be able to decrypt the obtained data.

\subsection{Range queries}

\begin{figure}
	\begin{center}
        \subfloat[Query searching for objects with a property $weight \ge 16$, $limit=4$ (takes 4 requests w/o cache)]{\label{fig:range-query-iter}\includegraphics[width=0.47\columnwidth]{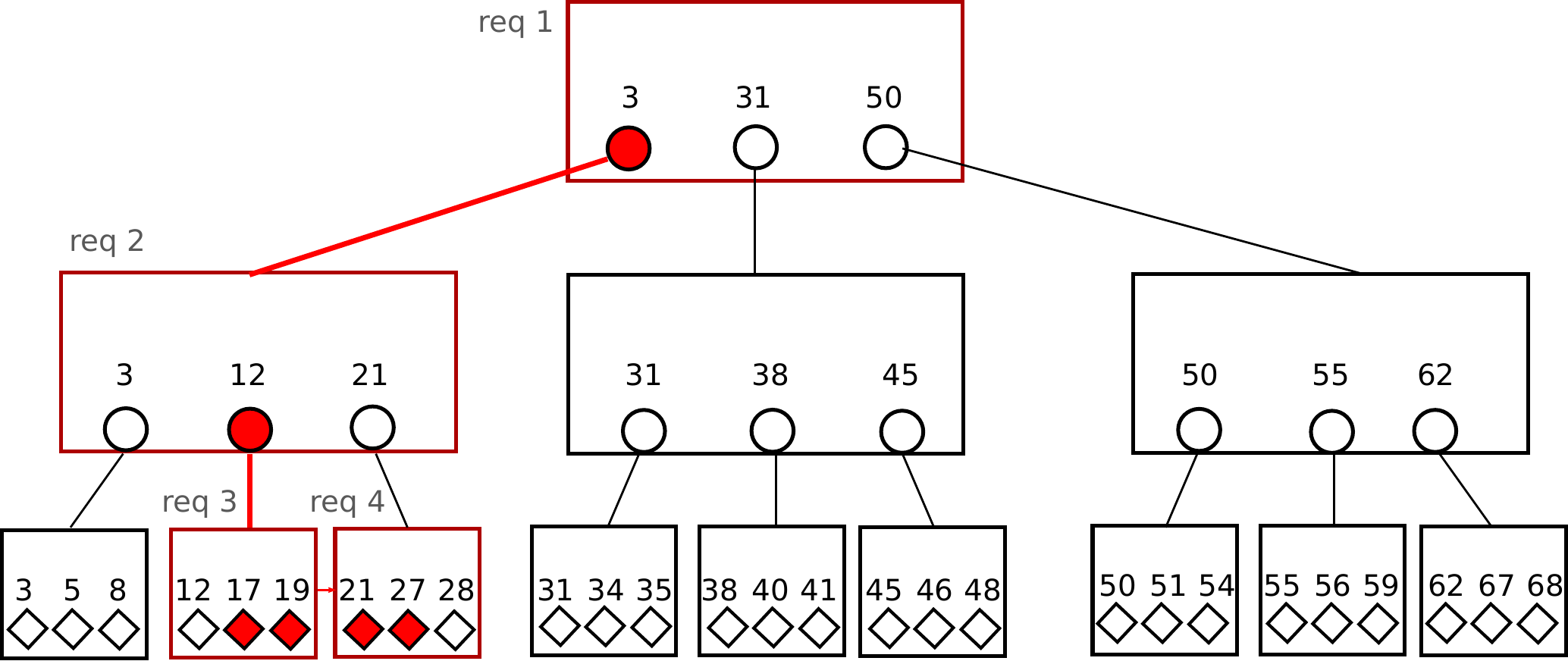}}
        \qquad
        \subfloat[Query which fetches \emph{all} objects  with $16 \le weight \le 27$ (takes 3 requests w/o cache)]{\label{fig:range-query-star}\includegraphics[width=0.47\columnwidth]{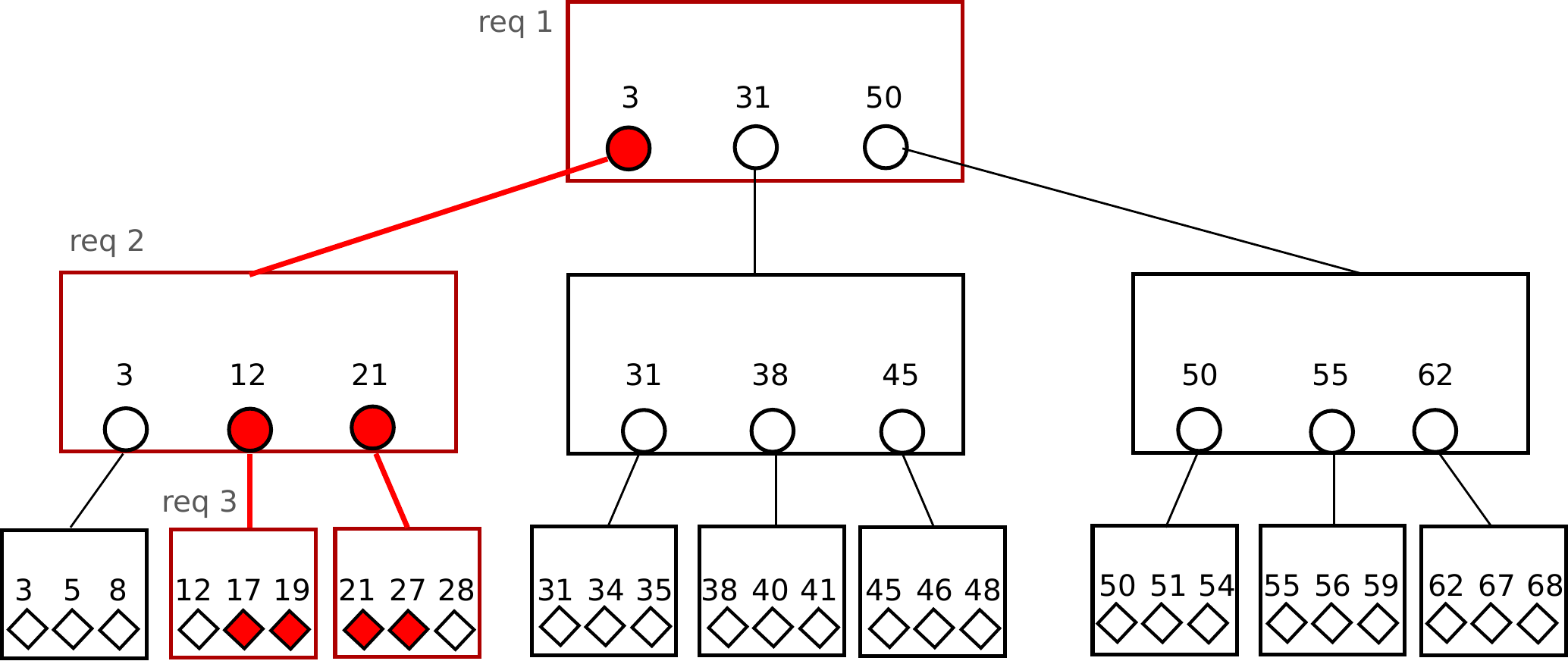}}
	\end{center}
	\caption{Range queries}
	\label{fig:range-query}
\end{figure}

When data is organized in B-Trees, range queries are trivial.
Take an example having records, \emph{Record}, with an integer property, \emph{weight}.
Data pointers to \emph{Record} objects are placed in a B-Tree sorted by \emph{weight}.

Two different types of range queries could be performed.
One is when we want a small subset of data in the beginning of the range (limit query).
In this case, we find a pointer to the beginning of the range, as in~\figref{fig:range-query-iter}, then we incrementally fetch subsequent buckets if the range occupies more than one, then bulk-fetch the objects by their pointers.

The other case is when we want to get \emph{all} objects in a range (select *).
We download the subsection of the B-Tree matching the range query level-by-level in a logarithmic number of steps, as in~\figref{fig:range-query-star}.
To do that, we start with the root bucket.
Then, we download all the child buckets which match the range at once, repeating until we get to the leaf nodes.
After that we (optionally) bulk-fetch all the objects which match the range query at once.
This is done in a way similar to prefetching all trees (Section~\ref{sec:bulk-fetching}).
Selecting all objects in a range reveals the approximate number of objects in this range to the server (the range and field names remain secret).

\subsection{Complex queries (multi-keyword search, multiple conditions)}

So far we have described simple queries.
But queries may be more complex and contain multiple conditions.
Depending on the number of objects matching each condition, and our desired security properties, we can select between two different approaches.
Making a query with an ``or'' condition requires simply zip-joining two sorted datasets, so we do not consider that to be problematic.
Instead, let's consider a query where we select objects matching the condition $(v_1 = a) \,\&\, (v_2 = b)$ and ordered by $v_3$.

\subsubsection{Prefetch approach}
\label{sec:prefetch}

In case the number of items with $v_1=a$ is small, it could be efficient to download the entire subset of matching object IDs.

We can use the following indexes for this kind of query:
\begin{align*}
    & \mbox{BTree}(v_1 \rightarrow \mbox{TreeSet}(ids)),\\
    & \mbox{BTree}(v_2 \rightarrow \mbox{TreeSet}(ids)),\\
    & \mbox{BTree}(id \rightarrow v_3).
\end{align*}

First, we estimate which condition has the smallest number of matching elements.
We do so by fetching contours of the B-Trees (corresponding to smallest and largest elements of the range). Since we know the average size of a bucket, we can determine the height of the tree and approximate distance between the smallest and largest elements.
This only requires a number of requests between client and server equal to the height of the tree, $H$.

We prefetch a TreeSet for the most lightweight condition (Sec.~\ref{sec:bulk-fetching}).
Then we bulk-search~(Sec.~\ref{sec:parallel-traversal}) these IDs in the larger TreeSet (and if we do not find an ID in the leaf node, we drop it).

After that, we need to sort the small subset of fetched IDs by $v_3$.
We do a parallel traversal of the third B-Tree and find which value of $v_3$ corresponds to each ID.

This approach reveals the number of elements matching condition $v_1=a$ to the server, although the condition itself remains unknown.

\subsubsection{Preorder approach}

The prefetch approach can be slow and reveal too much information to the server.
So, whenever possible, values are pre-ordered in the index.
It works in the following way.

The composite index to make a query selecting objects matching the condition $(v_1 = a) \,\&\, (v_2 = b)$ and ordered by $v_3$ is:
\begin{equation*}
    \mbox{BTree}((v_1, v_2) \rightarrow \mbox{BTree}(v_3 \rightarrow \mbox{TreeSet}(ids))).
\end{equation*}
For this query, we find a B-Tree (or multiple B-Trees) corresponding to the conditions and lazily traverse them in order of $v_3$.
This makes limit orders much more performant and doesn't leak information about the dataset size to the server.

\subsection{Incremental full-text search}
\label{sec:fulltext}

Full-text search with keywords, $t_1, t_2,\ldots, t_n$, requires ordering results by a net score or rank, which is determined according to a document's relevance to a given keyword.
One rarely needs all the matching documents at once, and the result is almost always sorted by the score.
This makes it possible to do the search in an incremental, pre-ordered manner, despite the score depending on the query.

Okapi BM25~\cite{wiki:OkapiBM25} is considered to be a state-of-art ranking function.
However, when it comes to scalability, it doesn't work well because term frequency (TF) depends on both the length of a document
and the mean length of documents in the database.
This makes it impossible to pre-compute a score even for one keyword, $t$, without having to fetch the individual numbers of occurrences of $t$
in all the documents where it appears.
Instead, we use Lucene's practical scoring function~\cite{lucene-practical-scoring}.

In a simple form, the score is determined as follows.
For a term, $t$, and document, $D$:
$$TF(t, D) = \sqrt{f(t, D)},$$
where $f(t, D)$~is the number of occurrences of $t$ in $D$. And the inverse document frequency (IDF) is:
$$IDF(t) = 1 + \ln\left( \frac{N_{docs}}{N_{docs}(t) + 1} \right),$$
where $N_{docs}$ is the total number of text documents indexed and $N_{docs}(t)$ is the number of documents which contain term $t$.
The values of $N_{docs}$ and $N_{docs}(t)$ are stored in an encrypted form and updated when new documents are added.
For easy conflict resolution, these values can be encrypted using additively homomorphic encryption (Pailier or exponential ElGamal).
The score of a document for one term, $t$, will be:
$$s(t, D) = \frac{TF(t, D) \cdot IDF(t)^2}{\sqrt{|D|}},$$
where $|D|$ is the total number of unique terms in document $D$.
$TF(t,D)/\sqrt{|D|}$ is calculated at indexing time, while $IDF$ is calculated at query time since it doesn't depend on $D$.
Thus, for one-keyword queries, documents are pre-ordered by relevancy.

Total document weight is defined as:
$$s(D) = \frac{\sum_{i=0}^n s(t_i, D)}{\sqrt{\sum IDF(t_i)^2}}.$$
This results in summing up the scores, $TF(t,D)/\sqrt{|D|}$, by which document references are pre-ordered (in descending order)
with weights, $IDF^2/\sqrt{\sum IDF}$, calculated at query time.

In order to produce document IDs sorted by $s$ incrementally, we first fetch tuples $(s_j(t_i, D), docid_j)$ with the highest values of $s$ for all the query terms, $t_i$.
Then we calculate possible ranges of document scores, given our knowledge of the highest known scores for terms $t_i$ and scores for particular $docids$ which we've read.
For example, a range of scores for a document, $D$, appearing in a sorted data structure would be:
$$\min(s(D)) = \sum_{\forall i~s_j(t_i, D)\in\text{cache}} s_j(t_i, D),\qquad
\max{s(D)} = \min(s(D)) + \sum_{\forall i~s_j(t_i, D)\notin\text{cache}} \min_{\in\text{cache}}\left(s(t_i, *)\right).$$
If the known documents which haven't yet been returned don't overlap by their range of possible scores when sorted by $\min(s)$,
it is fine to grab documents from the cache.
If the ranges of possible scores overlap, we don't have enough information and we read more document scores of the keyword for which the difference between scores is currently the highest.

This algorithm doesn't reveal any dataset sizes when we're interested in only a fraction of documents, which is beneficial for both security and scalability.
It is efficient when document weights for different keywords are significantly different from each other.

\subsection{Related objects}

ZeroDB is based on ZODB, which uses object references instead of joins~\cite{zodb-references}
(similar to MongoDB's database references~\cite{mongo-db-references}).
In order to keep everything consistent, ZODB fires events at commit time~\cite{zope-events} to keep back-references and indexes up to date.
In ZeroDB (as in ZODB), we can create indexes on attributes of related objects, much as we create normal indexes.

\subsection{Optimizations specific for remote client}

In most cases, all the query logic in ZeroDB happens client-side and the client and storage are separated by a network channel with high latency.
We use two primitives specific for this architecture.

\begin{figure}
	\begin{center}
        \subfloat[When a tree (or sub-tree) is small, it can be fully pre-fetched to the client]{\label{fig:fetch-tree}\includegraphics[width=0.47\columnwidth]{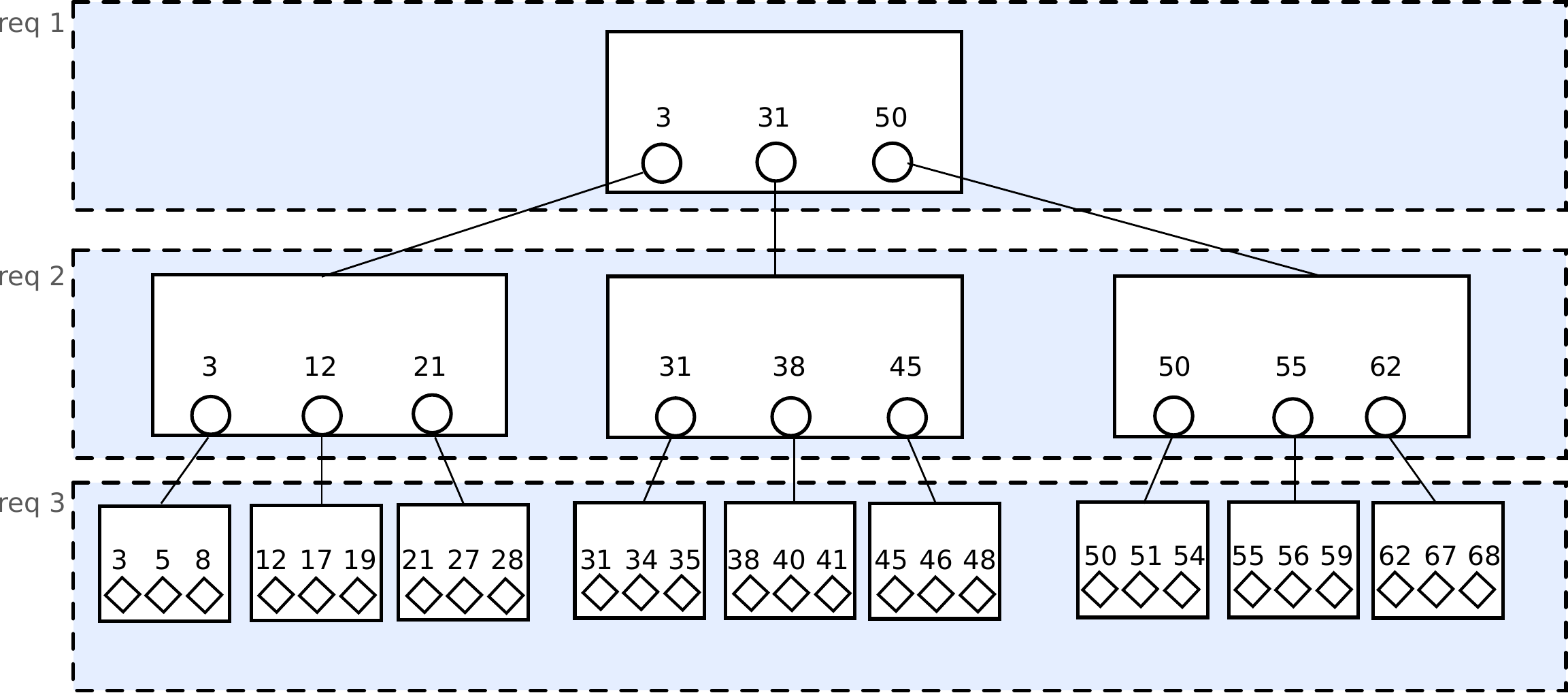}}
        \qquad
        \subfloat[When multiple branches of the tree needs reading/updating, tree traversal can be done in parallel]{\label{fig:parallel-traversal}\includegraphics[width=0.47\columnwidth]{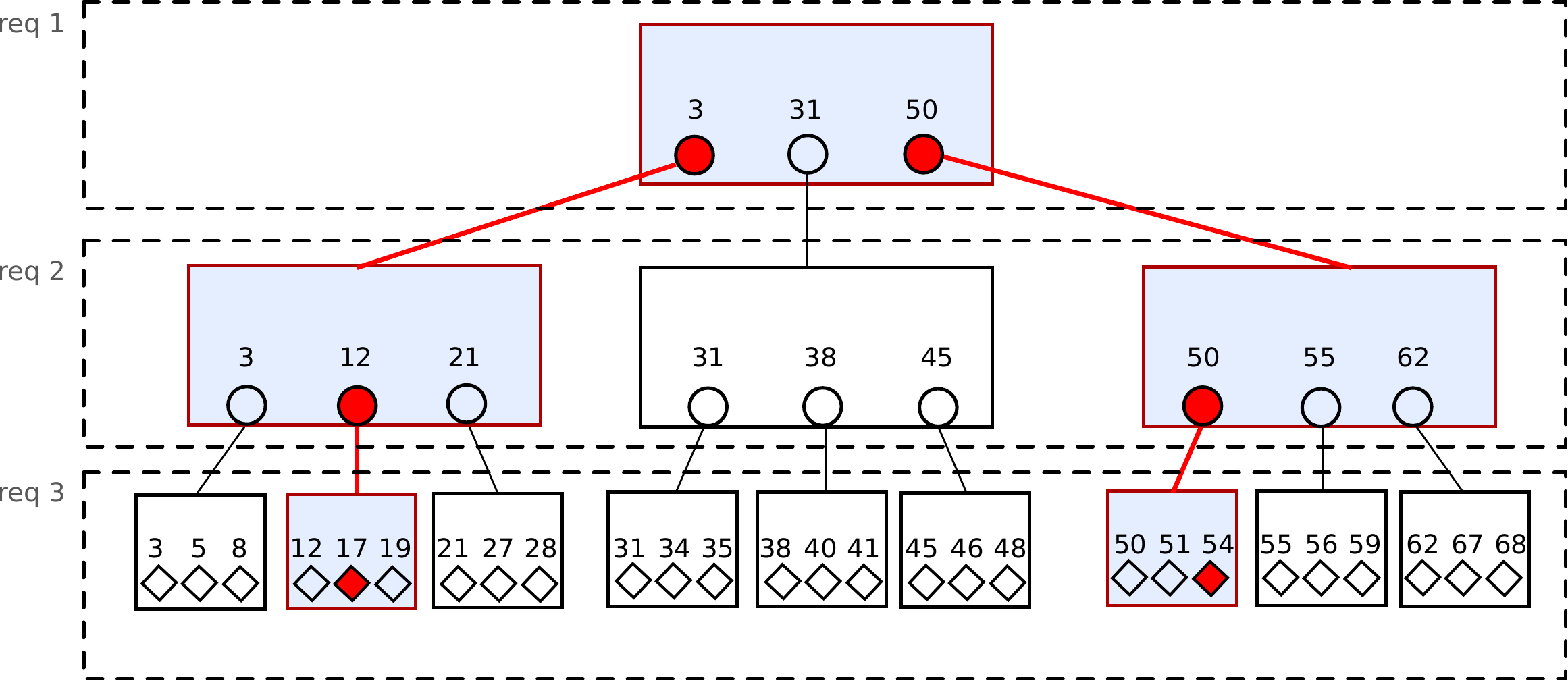}}
	\end{center}
    \caption{ZeroDB-specific optimizations for working with B-Trees includ pre-fetching a tree or finding multiple values in a number of steps equal to the tree height}
	\label{fig:tree-traversal-optimizations}
\end{figure}

\subsubsection{Bulk-fetching small (sub)trees}
\label{sec:bulk-fetching}

When a tree (or sub-tree) is small enough, it can be more performant to fetch the entire tree to the client.
We can do that in a logarithmic number of steps, rather than bucket-by-bucket, as we would do with an index stored on a local hard drive.

To do so, we fetch the root first, as in~\figref{fig:fetch-tree}.
The client decrypts the root bucket and fetches all of its children in a second query.
It unpacks all the child nodes and learns the IDs of their children.
This process repeats until the entire B-Tree is fetched.
Thus, the number of queries is equal to the height of the B-Tree, $H$, proportional to the logarithm of its size, $\log{S_{\mbox{ix}}}$.

When we fetch a tree, we implicitly show the number of objects it contains, $S_{\mbox{ix}}$, to the server.
Based on the size of the read data, the observer would be able to infer a number close to $S_{\mbox{ix}}$ and associate it with the bucket IDs just read.
Therefore, this technique should be used rarely, and in combination with reading other data at the same time or with oblivious RAM~\cite{path-oram,burst-oram,oram-multicloud,ods-wang-2014,practical-oram}.
The latter would prevent the server from learning bucket IDs.

\subsubsection{Parallel tree traversal}
\label{sec:parallel-traversal}

Often, queries involve several values of the same field (or different fields).
For example, indexing a new document containing $100$ words can be an expensive operation since it would involve $100\,H$ requests to the server.

In order to fix that, we traverse the B-Tree in parallel, as in~\figref{fig:parallel-traversal}.
We first fetch the root of the tree.
Then, we fetch only those child-buckets which are relevant to the values in our query, repeating as necessary.
This way, we do tree traversal for all the necessary values in $H$ steps.

When we do parallel traversal, the server would be able to see access patterns but it would not be able to distinguish access patterns belonging to each of the values individually.

\section{Sharing data}

It is often advantageous to share a dataset (for data analysis by third parties,
to give temporary accesss to employees or regulators, etc.).

When there are multiple users of the same encrypted dataset, there are serious shortcomings with sharing the encryption key with all parties.
Since the architecture of ZeroDB allows using any encryption algorithm, we can use proxy re-encryption~\cite{afgh,libert2011unidirectional}
or delta-keys~\cite{delta-keys,mylar} to share data with other users (third parties) of the same dataset.
These algorithms allow the third party to hold an encryption key other than that used to encrypt the data,
yet still query the data, until the data owner revokes their access.
By combining these sharing algorithms with ZeroDB's query functionality, we can granularly share subsets of data. 

\subsection{Proxy re-encryption}

Sharing an entire dataset is done as follows.
Suppose user $A$ owns the dataset and wants to share it with user $B$.
In the case of proxy re-encryption algorithms~\cite{afgh,libert2011unidirectional}, user $A$ has a key pair, $priv_a/pub_a$, and user $B$ has a key pair, $priv_b/pub_b$.
All objects and index buckets are encrypted with random, distinct content-encryption keys, $cek_i$, using a block cipher algorithm.
The $cek$ is encrypted with a key encrypting public key, $pub_a$.
So, we store $e_a(cek_i) = \mbox{encrypt}(pub_a, cek_i)$ and $c_i(obj) = \mbox{encrypt}(cek_i, data)$.

When user $A$ wants to give user $B$ access to the entire dataset and indexes, he constructs a re-encryption key, $r_{ab}$, such that
the server can transform the content-encryption key as $e_b(cek_i) = \mbox{transform}(r_{ab}, e_a(cek_i))$.
The re-encryption key, $r_{ab}$, is given to the server, along with an optional revocation time.
While the server has this re-encryption key, it transforms CEKs of encrypted objects, $e_a(cek_i) \rightarrow e_b(cek_i)$, as client $B$ requests them.
The server can revoke the key by removing $r_{ab}$.
Thus, while $r_{ab}$ is valid and present on the server, client $B$ is able to do tree traversals in the same manner as client $A$ (Sec.~\ref{sec:query-protocol}).

\subsection{Granular access permissions}

Let's first consider granular sharing of data in a scheme where $A$ shows encryption keys to $B$ and shares $M$ objects selected by a range query.
It is possible to share data granularly in such a way that $A$ has to only send $B$ a number of keys proportional to $\log{M}$ rather than $M$.

\begin{figure}
    \begin{center}
        \includegraphics[width=0.7\columnwidth]{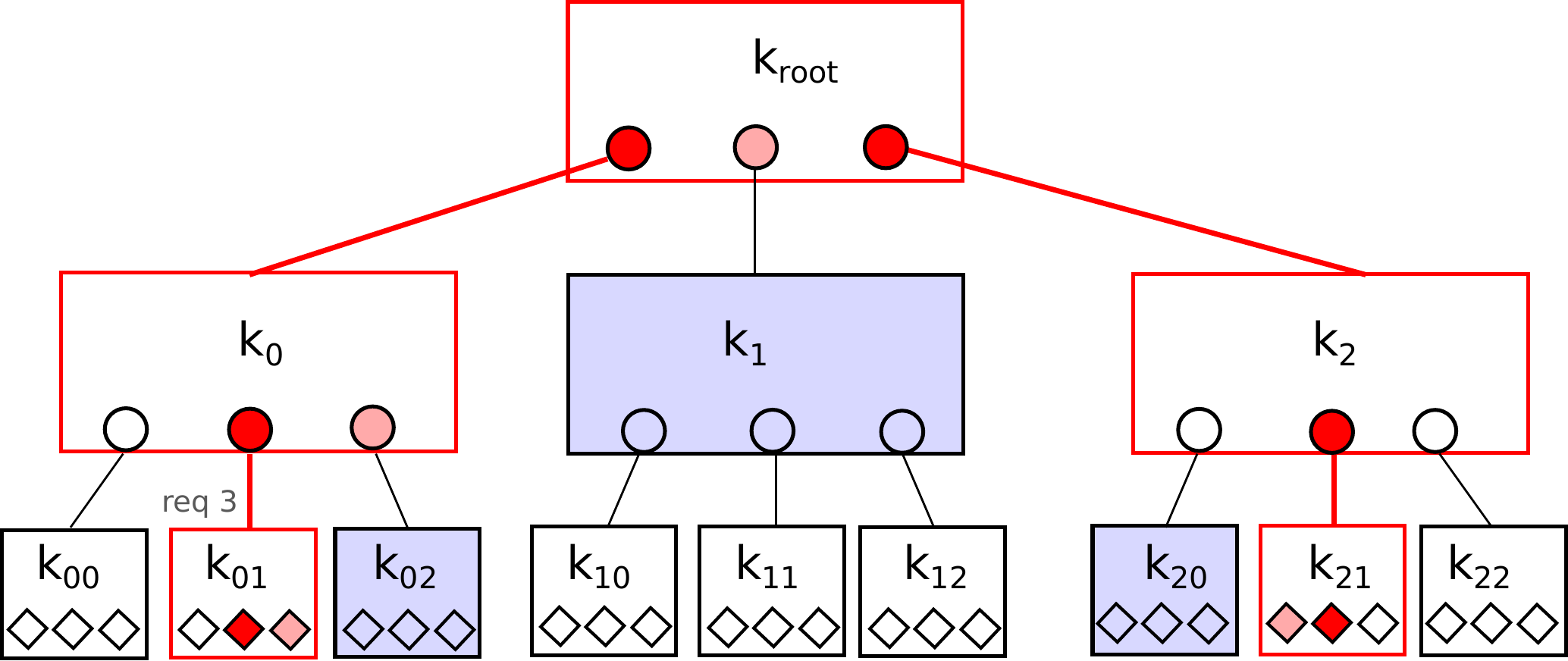}
    \end{center}
    \caption{
        B-Tree structure which supports selecting keys for any range of data.
        The hierarchy of keys is the same as its corresponding hierarchy of buckets.
        Giving keys from inside the contour would give access to objects only within the range.
    }
    \label{fig:keytree}
\end{figure}

Keys, $k^a$, must form a tree hierarchy, as in~\figref{fig:keytree}.
Each node of a B-Tree has a key associated with it.
Keys for child nodes $k_{*i}$ are derived from a key for the parent node $k_*$.
The keys can be derived as $k_{*i} = \mbox{SHA256}(k_* + i)$.
Client $A$ traverses the tree along the beginning and end of the range.
It shares keys for those buckets with $B$, giving $B$ the ability to do tree traversal within this range (the buckets with blue background in~\figref{fig:keytree}).
However, there may be buckets which are partly within the range.
Client $A$ passes key-reference pairs for buckets which belong to the contour of the range (depicted as red diamonds).
There should be no more than $2\,s_b$ such pairs, where $s_b$ is the maximum capacity of a bucket.
The set of keys giving access to the in-range fraction of the index and the in-range contents of the buckets from the contour of the range are given to $B$, encrypted with his public key.

The described sharing mechanism suffers from the fact that $B$ is given all the information to decrypt objects within the range at once.
In order to make this access revokable, we need to use proxy re-encryption.
In this scenario, client $A$ gives hierarchical rights to re-encrypt the data to the server rather than to client $B$ directly.
That is, the server should only be able to re-encrypt objects within the range, given a set of re-encryption keys, $r_{ab}$.
It is possible to do this using a conditional proxy re-encryption scheme (C-PRE)~\cite{conditional-pre-2009,unidirectional-pre-2010,conditional-pre-2014,hierarchical-pre}.

Usually, such proxy re-encryption schemes support giving a product of multiple equality conditions.
The conditions which we apply would be object IDs in the bucket hierarchy, starting from the top.
For example, C-PRE encryption of a bucket in the second layer of the tree, $b_1$, would give the server the ability to derive re-encryption keys for all of its children, $b_{1*}$.
This gives the server the ability to re-encrypt all buckets within the range specified by $A$, while only requiring $A$ to calculate a number of keys proportional to the $\log$ of the number of records shared.

\section{Security analysis}

ZeroDB is agnostic to the encryption algorithm used.
By default, we use authenticated strong encryption ({AES-256} in GCM mode).
Nothing is decrypted on the server.

Using deterministic or order-preserving encryption visible to the server can be detrimental to the security of the database in certain cases~\cite{cryptdb-hacked}.
However, even traversal of B-Trees allows an attacker to collect access statistics and infer the data in the database~\cite{access-pattern-attack}.
We provide a security analysis and an example of the number of queries necessary for such an attack.

\subsection{Threat model}

We adopt a threat model similar to the one used for the security analysis of CipherBase~\cite{cipherbase}.
We introduce a \emph{resident adversary}, who has unlimited observational power and can view the contents of memory and disk at every moment of time.
An \emph{instant adversary} can observe only a snapshot of memory and disk at one particular moment in time.
We assume that both resident and instant adversaries are passive (honest-but-curious) and do not return false results to the client.
In the future, we may consider active adversaries and modify our protocol accordingly.

\subsection{Data confidentiality}

\begin{figure}
	\begin{center}
        \includegraphics[width=0.7\columnwidth]{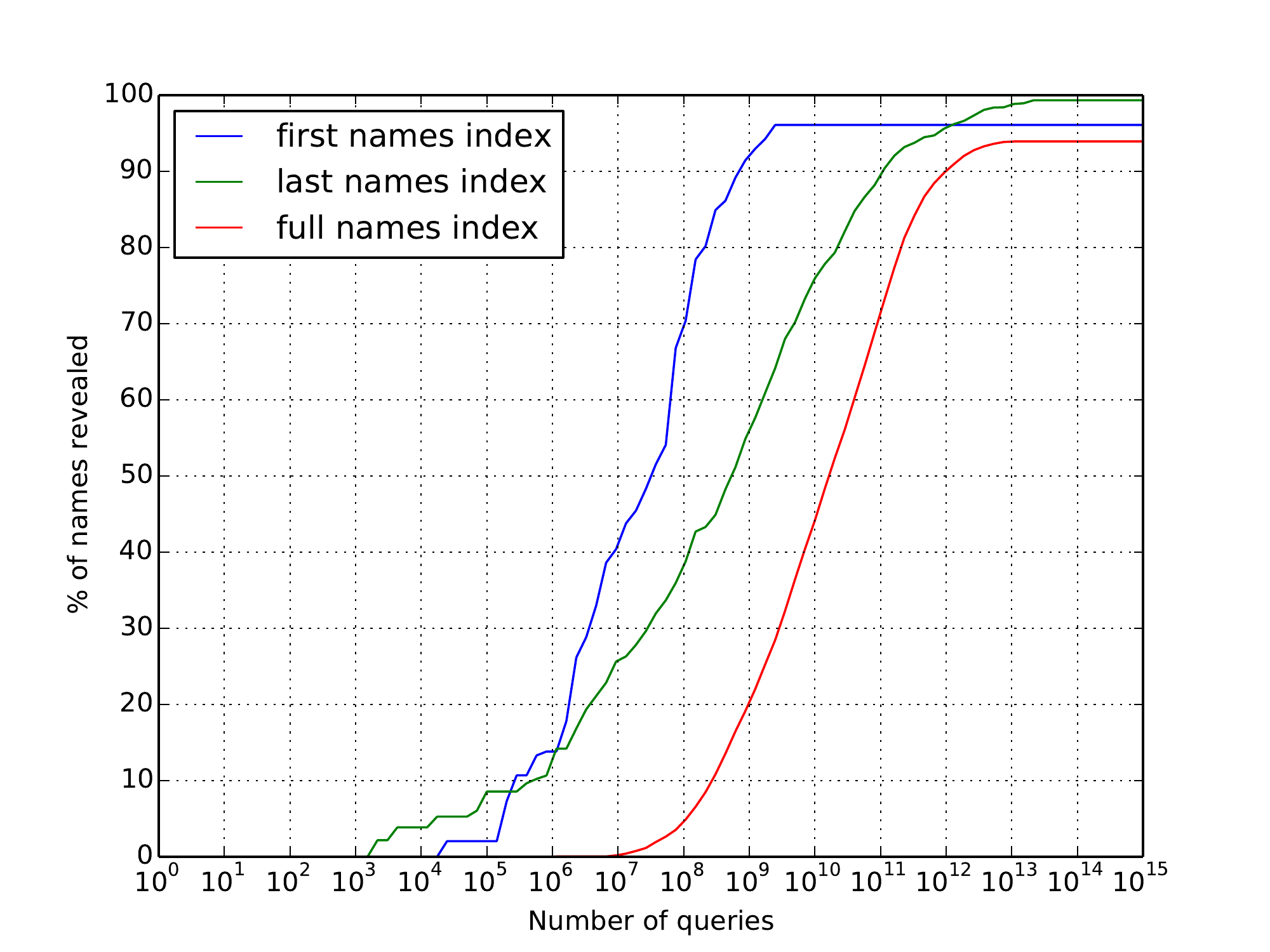}
	\end{center}
    \caption{
        Fraction of queries where last name can be inferred by a passive observer depending on the number of queries made to the database.
        The fraction of names able to be inferred depends on whether we index first and last names separately or together.
        In CryptDB, this information is available to an attacker immediately~\cite{cryptdb}.
    }
	\label{fig:surname-attack}
\end{figure}

In ZeroDB, all contents of the database and its indexes are stored as encrypted objects.
Therefore, an instant adversary doesn't learn anything except for database size (including the number of fields).
So, ZeroDB provides \emph{semantic security} against an instant adversary.

A resident adversary, on the other hand, is able to observe access patterns and collect statistics of those over time.
So, we provide \emph{operational security} against a resident adversary.

First, we consider search queries which only require tree traversal.
For example, limit orders for queries which do not involve pulling a large subset of the index to the client.
For simplicity, let's assume that all we have in a database index is a B-Tree of customer surnames.
A resident adversary is able to acquire access statistics of the index leaf nodes when searches of, say, $10$ people with a given last name are performed.
However, the amount of statistics should be large enough to distinguish different records in the database.
If we assume the probability of a given surname appearing in a given query is the same as in the U.S. Census 2000 data~\cite{us-census-surnames}
and that keywords are distributed across queries according to a Poisson distribution,
an adversary needs to observe two keywords at least $N_1$ and $N_2$ times, such that $|N_1 - N_2| > \sqrt{N_1 + N_2}$.
The fraction of last names which can be revealed by a resident adversary from a query, given some number of queries, is shown on~\figref{fig:surname-attack}.
For example, the surnames in $10\%$ of new queries can be inferred by a resident adversary after observing a million queries (although surnames not touched by queries are not revealed).
However, if we index full names as one field (first names are taken from UK Office of national statistics~\cite{uk-first-names}),
we would need to do $300$ millions queries to infer $10\%$ of full names in new queries.
This assumes that an attacker is aware of the statistics of data in the database, and nothing other than names are queried.
So, these estimates provide a lower bound for the number of queries required to infer useful information.

Unlike CipherBase~\cite{cipherbase}, our approach doesn't leak information about the order of the elements.
By observing access patterns, the server can only figure out which buckets are leaf nodes and which are branch nodes,
but not whether the child node is located to the left or right of the parent node.

In cases where we need to prefetch a large chunk of a B-Tree~(Sec.~\ref{sec:prefetch}), we immediately reveal information about the size of the sub-dataset to the server.
Of course, this is undesirable.
To avoid this, we can download adjacent records in the same query, so that the downloaded size is always approximately constant and equal to that of the largest sub-dataset.
This results in similar security guarantees to those for tree traversal.
In order for this to work we need to unfold nested trees into one tree~(Sec.~\ref{sec:unfold-trees}).
The way we perform full-text searches~(Sec.~\ref{sec:fulltext}) doesn't reveal the numbers of documents per keyword since it's done in incremental manner.

We can significantly improve security guarantees by using ORAM techniques, which allow us to hide access patterns by writing back data after reading it.
Several algorithms are available, though the most promising ones are Ring ORAM~\cite{practical-oram} with XOR technique~\cite{burst-oram}
and oblivious data structures~\cite{ods-wang-2014}.
Still, these require writing back to the database about 5-20 times the amount of data read (although these numbers improve every year).
It is also possible that hiding access patterns in a multi-cloud or fully decentralized arrangement has quite attractive performance properties~\cite{oram-multicloud},
which is especially interesting in light of decentralized storage layers such as Storj.io~\cite{storj} and IPFS~\cite{ipfs}.
In fact, Enigma~\cite{enigma} is an early exploration of using a blockchain to guarantee privacy and fault-tolerance.

\section{Performance}

\subsection{Optimizing bucket size}

ZeroDB requires several interactions between client and server in order to return query results.
Thus, both connection bandwidth, $b$, and time of round-trip between client and server, $\tau$, determine the performance of queries.
Moreover, there is an optimal bucket size, $s_b$, for any given parameters.

Suppose we have a B-Tree index of size $s_i$ bytes.
The size of one record in the bucket is $s_r$ bytes.
Apart from encryption we can apply compression.
The compression rate is defined as the ratio of initial bucket size to final (compressed + encrypted) bucket size is $c$.
The number of requests necessary to perform a query with an empty cache is equal to the height of B-Tree, $h$, though we also consider the case where we cache up to $s_c$ bytes.

A good approximation for B-Tree height is~\cite{wiki:b-tree}:
$$h = \log_m n,$$
where $m = s_b/s_r$ is the number of records in one bucket and $n = s_i/s_r$ is the total number of records in the index,
assuming the best and the worst case heights are close to each other.
Thus:
$$h \approx \frac{\ln s_i - \ln s_r}{\ln s_b - \ln s_r}.$$

When caching, we cache predominantly the top of the B-Tree with an average cached height, $h_c$:
$$h_c \approx \frac{\ln s_c - \ln s_r}{\ln s_b - \ln s_r}.$$
The average number of interactions between client and server per query will be $k = h - h_c$.
Thus, the time which this interaction takes is:
$$\Delta t = k\tau + k\frac{s_b}{cb} =
\left(h - h_c \right) \left( \tau + \frac{s_b}{cb} \right) =
\frac{\ln s_i - \ln s_c}{\ln s_b - \ln s_r} \left( \tau + \frac{s_b}{cb} \right).$$

We make queries as fast as possible (i.e.\ minimize time $\Delta t$ necessary for a query to complete) by seeking for the optimal bucket size, $s_b$.
The largest possible bucket size is equal to the size of the index and is obviously not efficient for sufficiently large indexes (it takes too long to download the index).
On the other hand, when bucket size is too small, the number of requests is too large and query performance suffers due to latency between client and server.
An optimal compromise can be determined by equalizing the first derivative to zero:
$$\frac{\partial \Delta t}{\partial s_b} = 0,$$
which yields the optimal bucket size:
$$s_b = \frac{cb\tau}{W\left( \frac{cb\tau}{e s_r}\right)},$$
where $W$ is the real part of Lambert W function~\cite{wiki:lambert}
and $e=2.718281828\ldots$.

In ZeroDB, the typical size of a record in a branch bucket is $s_r = 29$ bytes, with a zlib compression rate of $c=3$.
For a typical network bandwidth of $b=10^6$ (1~MB/s) and roundtrip time of $\tau=0.05~\mbox{s}$, the optimal bucket size is $s_b=10$~kB when compression is not used, or $8$~kB when compression is used.
For a typical cache size of $5$~MB and index size of $50$~GB,
an average query takes $t=.07$~s and requires network traffic of $11$~kB of compressed data (without caching that would take $0.17$~s and network traffic of $24$~kB of compressed data), reducing the average number of roundtrips to $1.3$.
We can finetune the optimal bucket size on the fly as the connection speed improves over time.

When we need to traverse several indexes at once, we bunch requests for several indexes in one query.
For example, the first request for indexes $A$ and $B$ would be ``read the roots of indexes $A$ and $B$ (by object IDs of the roots)''.
In the past, that would make heavier queries impractical, but the growth in internet bandwidth according to Nielsen's law~\cite{nielsen-law} alleviates this.

\subsection{Unfolding trees with subtrees into one tree}
\label{sec:unfold-trees}

By default, we often use nested B-Trees of TreeSets.
While that yields performance benefits in traditional databases, it is not ideal when working with large latencies.
And, more importantly, it has security drawbacks, since we cannot easily read adjacent records to hide the size of the dataset from the server.

Consider a B-Tree linking word IDs, $wid$, with TreeSets of object IDs, $oid$.
We can instead use a treeset of tuples (wid,~oid) which will be unique.
The tree will have approximately the same height for all $wid$s.
Also, it is easy to pre-set the approximate size of the dataset to download if we don't want to show how many $oid$s we have per $wid$.

\section{Cryptographic primitives used}

Even though the server never decrypts the data, we use encryption in transit (using SSL) in order to prevent third parties from analyzing access patterns.
The client authenticates with an X.509 certificate or a self-signed certificate where the private key is derived from a passphrase.
When a passphrase is used, the scrypt algorithm is used to derive a private key from the passphrase. Information about the salt is stored server-side and is given before establishing an authenticated encrypted connection.

For symmetric encryption, we use AES256 in GCM mode.
The symmetric key is derived from either the passphrase, the private key in an X.509 certificate, or provided by a key management service (KMS).

For proxy re-encryption, we use the AFGH algorithm~\cite{afgh} to encrypt a random content key for an AES256 block cipher in GCM mode.

\section{Acknowledgements}

We thank Prof. David Evans from the University of Virginia and Prof. Nigel Smart from the University of Bristol for reviewing this white paper
and giving advice for further improvements in the security and performance of ZeroDB.
We also thank Prof. Nickolai Zeldovich, Christopher Fletcher, Albert Kwon and Ling Ren from MIT for their advice at earlier stages of ZeroDB's development.
Additionally, we thank Dr. Raluca Ada Popa from UC Berkeley for fruitful discussions about ways to offload complicated query processing to the server
in a secure manner.

\bibliography{zerodb}

%merlin.mbs apsrev4-1.bst 2010-07-25 4.21a (PWD, AO, DPC) hacked
%Control: key (0)
%Control: author (0) dotless jnrlst
%Control: editor formatted (1) identically to author
%Control: production of article title (0) allowed
%Control: page (1) range
%Control: year (0) verbatim
%Control: production of eprint (0) enabled
\begin{thebibliography}{35}%
\makeatletter
\providecommand \@ifxundefined [1]{%
 \@ifx{#1\undefined}
}%
\providecommand \@ifnum [1]{%
 \ifnum #1\expandafter \@firstoftwo
 \else \expandafter \@secondoftwo
 \fi
}%
\providecommand \@ifx [1]{%
 \ifx #1\expandafter \@firstoftwo
 \else \expandafter \@secondoftwo
 \fi
}%
\providecommand \natexlab [1]{#1}%
\providecommand \enquote  [1]{``#1''}%
\providecommand \bibnamefont  [1]{#1}%
\providecommand \bibfnamefont [1]{#1}%
\providecommand \citenamefont [1]{#1}%
\providecommand \href@noop [0]{\@secondoftwo}%
\providecommand \href [0]{\begingroup \@sanitize@url \@href}%
\providecommand \@href[1]{\@@startlink{#1}\@@href}%
\providecommand \@@href[1]{\endgroup#1\@@endlink}%
\providecommand \@sanitize@url [0]{\catcode `\\12\catcode `\$12\catcode
  `\&12\catcode `\#12\catcode `\^12\catcode `\_12\catcode `\%12\relax}%
\providecommand \@@startlink[1]{}%
\providecommand \@@endlink[0]{}%
\providecommand \url  [0]{\begingroup\@sanitize@url \@url }%
\providecommand \@url [1]{\endgroup\@href {#1}{\urlprefix }}%
\providecommand \urlprefix  [0]{URL }%
\providecommand \Eprint [0]{\href }%
\providecommand \doibase [0]{http://dx.doi.org/}%
\providecommand \selectlanguage [0]{\@gobble}%
\providecommand \bibinfo  [0]{\@secondoftwo}%
\providecommand \bibfield  [0]{\@secondoftwo}%
\providecommand \translation [1]{[#1]}%
\providecommand \BibitemOpen [0]{}%
\providecommand \bibitemStop [0]{}%
\providecommand \bibitemNoStop [0]{.\EOS\space}%
\providecommand \EOS [0]{\spacefactor3000\relax}%
\providecommand \BibitemShut  [1]{\csname bibitem#1\endcsname}%
\let\auto@bib@innerbib\@empty
%</preamble>
\bibitem [{\citenamefont {Arasu}\ \emph
  {et~al.}(2014{\natexlab{a}})\citenamefont {Arasu}, \citenamefont {Eguro},
  \citenamefont {Kaushik},\ and\ \citenamefont {Ramamurthy}}]{tutorial}%
  \BibitemOpen
  \bibfield  {author} {\bibinfo {author} {\bibfnamefont {Arvind}\ \bibnamefont
  {Arasu}}, \bibinfo {author} {\bibfnamefont {Ken}\ \bibnamefont {Eguro}},
  \bibinfo {author} {\bibfnamefont {Raghav}\ \bibnamefont {Kaushik}}, \ and\
  \bibinfo {author} {\bibfnamefont {Ravishankar}\ \bibnamefont {Ramamurthy}},\
  }\bibfield  {title} {\enquote {\bibinfo {title} {Querying encrypted data},}\
  }in\ \href {\doibase 10.1145/2588555.2588893} {\emph {\bibinfo {booktitle}
  {Proceedings of the 2014 ACM SIGMOD International Conference on Management of
  Data}}},\ \bibinfo {series and number} {SIGMOD '14}\ (\bibinfo  {publisher}
  {ACM},\ \bibinfo {address} {New York, NY, USA},\ \bibinfo {year} {2014})\
  pp.\ \bibinfo {pages} {1259--1261}\BibitemShut {NoStop}%
\bibitem [{\citenamefont {Gentry}(2010)}]{gentry}%
  \BibitemOpen
  \bibfield  {author} {\bibinfo {author} {\bibfnamefont {Craig}\ \bibnamefont
  {Gentry}},\ }\bibfield  {title} {\enquote {\bibinfo {title} {Computing
  arbitrary functions of encrypted data},}\ }\href {\doibase
  10.1145/1666420.1666444} {\bibfield  {journal} {\bibinfo  {journal} {Commun.
  ACM}\ }\textbf {\bibinfo {volume} {53}},\ \bibinfo {pages} {97--105}
  (\bibinfo {year} {2010})}\BibitemShut {NoStop}%
\bibitem [{\citenamefont {Popa}\ \emph {et~al.}(2011)\citenamefont {Popa},
  \citenamefont {Redfield}, \citenamefont {Zeldovich},\ and\ \citenamefont
  {Balakrishnan}}]{cryptdb}%
  \BibitemOpen
  \bibfield  {author} {\bibinfo {author} {\bibfnamefont {Raluca~Ada}\
  \bibnamefont {Popa}}, \bibinfo {author} {\bibfnamefont {Catherine M.~S.}\
  \bibnamefont {Redfield}}, \bibinfo {author} {\bibfnamefont {Nickolai}\
  \bibnamefont {Zeldovich}}, \ and\ \bibinfo {author} {\bibfnamefont {Hari}\
  \bibnamefont {Balakrishnan}},\ }\bibfield  {title} {\enquote {\bibinfo
  {title} {Cryptdb: Protecting confidentiality with encrypted query
  processing},}\ }in\ \href {\doibase 10.1145/2043556.2043566} {\emph {\bibinfo
  {booktitle} {Proceedings of the Twenty-Third ACM Symposium on Operating
  Systems Principles}}},\ \bibinfo {series and number} {SOSP '11}\ (\bibinfo
  {publisher} {ACM},\ \bibinfo {address} {New York, NY, USA},\ \bibinfo {year}
  {2011})\ pp.\ \bibinfo {pages} {85--100}\BibitemShut {NoStop}%
\bibitem [{\citenamefont {Kamara}\ \emph {et~al.}(2012)\citenamefont {Kamara},
  \citenamefont {Papamanthou},\ and\ \citenamefont {Roeder}}]{dynamicsse}%
  \BibitemOpen
  \bibfield  {author} {\bibinfo {author} {\bibfnamefont {Seny}\ \bibnamefont
  {Kamara}}, \bibinfo {author} {\bibfnamefont {Charalampos}\ \bibnamefont
  {Papamanthou}}, \ and\ \bibinfo {author} {\bibfnamefont {Tom}\ \bibnamefont
  {Roeder}},\ }\bibfield  {title} {\enquote {\bibinfo {title} {Dynamic
  searchable symmetric encryption},}\ }in\ \href {\doibase
  10.1145/2382196.2382298} {\emph {\bibinfo {booktitle} {Proceedings of the
  2012 ACM Conference on Computer and Communications Security}}},\ \bibinfo
  {series and number} {CCS '12}\ (\bibinfo  {publisher} {ACM},\ \bibinfo
  {address} {New York, NY, USA},\ \bibinfo {year} {2012})\ pp.\ \bibinfo
  {pages} {965--976}\BibitemShut {NoStop}%
\bibitem [{\citenamefont {Huang}\ \emph {et~al.}(2013)\citenamefont {Huang},
  \citenamefont {Katz},\ and\ \citenamefont {Evans}}]{mpc}%
  \BibitemOpen
  \bibfield  {author} {\bibinfo {author} {\bibfnamefont {Yan}\ \bibnamefont
  {Huang}}, \bibinfo {author} {\bibfnamefont {Jonathan}\ \bibnamefont {Katz}},
  \ and\ \bibinfo {author} {\bibfnamefont {Dave}\ \bibnamefont {Evans}},\
  }\href@noop {} {\enquote {\bibinfo {title} {Efficient secure two-party
  computation using symmetric cut-and-choose},}\ }\bibinfo {howpublished}
  {Cryptology ePrint Archive, Report 2013/081} (\bibinfo {year} {2013}),\
  \bibinfo {note} {\url{http://eprint.iacr.org/}}\BibitemShut {NoStop}%
\bibitem [{\citenamefont {Arasu}\ \emph
  {et~al.}(2014{\natexlab{b}})\citenamefont {Arasu}, \citenamefont {Eguro},
  \citenamefont {Joglekar}, \citenamefont {Kaushik}, \citenamefont {Kossmann},\
  and\ \citenamefont {Ramamurthy}}]{cipherbase}%
  \BibitemOpen
  \bibfield  {author} {\bibinfo {author} {\bibfnamefont {Arvind}\ \bibnamefont
  {Arasu}}, \bibinfo {author} {\bibfnamefont {Ken}\ \bibnamefont {Eguro}},
  \bibinfo {author} {\bibfnamefont {Manas}\ \bibnamefont {Joglekar}}, \bibinfo
  {author} {\bibfnamefont {Raghav}\ \bibnamefont {Kaushik}}, \bibinfo {author}
  {\bibfnamefont {Donald}\ \bibnamefont {Kossmann}}, \ and\ \bibinfo {author}
  {\bibfnamefont {Ravi}\ \bibnamefont {Ramamurthy}},\ }\href
  {http://research.microsoft.com/apps/pubs/default.aspx?id=226364} {\emph
  {\bibinfo {title} {Transaction Processing on Confidential Data using
  Cipherbase}}},\ \bibinfo {type} {Tech. Rep.}\ \bibinfo {number}
  {MSR-TR-2014-106}\ (\bibinfo {year} {2014})\BibitemShut {NoStop}%
\bibitem [{\citenamefont {Smart}\ \emph {et~al.}(2015)\citenamefont {Smart}
  \emph {et~al.}}]{smart}%
  \BibitemOpen
  \bibfield  {author} {\bibinfo {author} {\bibfnamefont {Nigel}\ \bibnamefont
  {Smart}} \emph {et~al.},\ }\href
  {https://www.cs.bris.ac.uk/~nigel/ECRYPT-MPC/Draft.pdf} {\emph {\bibinfo
  {title} {Future Directions in Computing on Encrypted Data (Draft)}}},\
  \bibinfo {type} {Tech. Rep.}\ (\bibinfo {year} {2015})\BibitemShut {NoStop}%
\bibitem [{\citenamefont {Wikipedia}(2016{\natexlab{a}})}]{wiki:OkapiBM25}%
  \BibitemOpen
  \bibfield  {author} {\bibinfo {author} {\bibnamefont {Wikipedia}},\ }\href
  {https://en.wikipedia.org/wiki/Okapi_BM25} {\enquote {\bibinfo {title} {Okapi
  {BM25}},}\ } (\bibinfo {year} {2016}{\natexlab{a}}),\ \bibinfo {note}
  {[Online; accessed 17-Feb-2016]}\BibitemShut {NoStop}%
\bibitem [{luc()}]{lucene-practical-scoring}%
  \BibitemOpen
  \href
  {https://www.elastic.co/guide/en/elasticsearch/guide/current/practical-scoring-function.html}
  {\enquote {\bibinfo {title} {Lucene's practical scoring function},}\
  }\bibinfo {note} {[Online; accessed 17-Feb-2016]}\BibitemShut {NoStop}%
\bibitem [{\citenamefont {Faassen}(2008)}]{zodb-references}%
  \BibitemOpen
  \bibfield  {author} {\bibinfo {author} {\bibfnamefont {Martijn}\ \bibnamefont
  {Faassen}},\ }\href
  {http://blog.startifact.com/posts/older/a-misconception-about-the-zodb.html}
  {\enquote {\bibinfo {title} {A misconception about the zodb},}\ } (\bibinfo
  {year} {2008})\BibitemShut {NoStop}%
\bibitem [{mon()}]{mongo-db-references}%
  \BibitemOpen
  \href {https://docs.mongodb.org/manual/reference/database-references/}
  {\enquote {\bibinfo {title} {Mongodb documentation: Database references},}\
  }\BibitemShut {NoStop}%
\bibitem [{zop()}]{zope-events}%
  \BibitemOpen
  \href {https://pypi.python.org/pypi/zope.lifecycleevent} {\enquote {\bibinfo
  {title} {zope.lifecycleevent},}\ }\BibitemShut {NoStop}%
\bibitem [{\citenamefont {Stefanov}\ \emph {et~al.}(2013)\citenamefont
  {Stefanov}, \citenamefont {van Dijk}, \citenamefont {Shi}, \citenamefont
  {Fletcher}, \citenamefont {Ren}, \citenamefont {Yu},\ and\ \citenamefont
  {Devadas}}]{path-oram}%
  \BibitemOpen
  \bibfield  {author} {\bibinfo {author} {\bibfnamefont {Emil}\ \bibnamefont
  {Stefanov}}, \bibinfo {author} {\bibfnamefont {Marten}\ \bibnamefont {van
  Dijk}}, \bibinfo {author} {\bibfnamefont {Elaine}\ \bibnamefont {Shi}},
  \bibinfo {author} {\bibfnamefont {Christopher}\ \bibnamefont {Fletcher}},
  \bibinfo {author} {\bibfnamefont {Ling}\ \bibnamefont {Ren}}, \bibinfo
  {author} {\bibfnamefont {Xiangyao}\ \bibnamefont {Yu}}, \ and\ \bibinfo
  {author} {\bibfnamefont {Srinivas}\ \bibnamefont {Devadas}},\ }\bibfield
  {title} {\enquote {\bibinfo {title} {Path {ORAM}: An extremely simple
  oblivious {RAM} protocol},}\ }in\ \href {\doibase 10.1145/2508859.2516660}
  {\emph {\bibinfo {booktitle} {Proceedings of the 2013 ACM SIGSAC Conference
  on Computer \& Communications Security}}},\ \bibinfo {series and number} {CCS
  '13}\ (\bibinfo  {publisher} {ACM},\ \bibinfo {address} {New York, NY, USA},\
  \bibinfo {year} {2013})\ pp.\ \bibinfo {pages} {299--310}\BibitemShut
  {NoStop}%
\bibitem [{\citenamefont {Dautrich}\ \emph {et~al.}(2014)\citenamefont
  {Dautrich}, \citenamefont {Stefanov},\ and\ \citenamefont
  {Shi}}]{burst-oram}%
  \BibitemOpen
  \bibfield  {author} {\bibinfo {author} {\bibfnamefont {Jonathan}\
  \bibnamefont {Dautrich}}, \bibinfo {author} {\bibfnamefont {Emil}\
  \bibnamefont {Stefanov}}, \ and\ \bibinfo {author} {\bibfnamefont {Elaine}\
  \bibnamefont {Shi}},\ }\bibfield  {title} {\enquote {\bibinfo {title} {Burst
  {ORAM}: Minimizing {ORAM} response times for bursty access patterns},}\ }in\
  \href
  {https://www.usenix.org/conference/usenixsecurity14/technical-sessions/presentation/dautrich}
  {\emph {\bibinfo {booktitle} {23rd USENIX Security Symposium (USENIX Security
  14)}}}\ (\bibinfo  {publisher} {USENIX Association},\ \bibinfo {address} {San
  Diego, CA},\ \bibinfo {year} {2014})\ pp.\ \bibinfo {pages}
  {749--764}\BibitemShut {NoStop}%
\bibitem [{\citenamefont {Stefanov}\ and\ \citenamefont
  {Shi}(2013)}]{oram-multicloud}%
  \BibitemOpen
  \bibfield  {author} {\bibinfo {author} {\bibfnamefont {Emil}\ \bibnamefont
  {Stefanov}}\ and\ \bibinfo {author} {\bibfnamefont {Elaine}\ \bibnamefont
  {Shi}},\ }\bibfield  {title} {\enquote {\bibinfo {title} {Multi-cloud
  oblivious storage},}\ }in\ \href {\doibase 10.1145/2508859.2516673} {\emph
  {\bibinfo {booktitle} {Proceedings of the 2013 ACM SIGSAC Conference on
  Computer \& Communications Security}}},\ \bibinfo {series and number} {CCS
  '13}\ (\bibinfo  {publisher} {ACM},\ \bibinfo {address} {New York, NY, USA},\
  \bibinfo {year} {2013})\ pp.\ \bibinfo {pages} {247--258}\BibitemShut
  {NoStop}%
\bibitem [{\citenamefont {Wang}\ \emph {et~al.}(2014)\citenamefont {Wang},
  \citenamefont {Nayak}, \citenamefont {Liu}, \citenamefont {Chan},
  \citenamefont {Shi}, \citenamefont {Stefanov},\ and\ \citenamefont
  {Huang}}]{ods-wang-2014}%
  \BibitemOpen
  \bibfield  {author} {\bibinfo {author} {\bibfnamefont {Xiao~Shaun}\
  \bibnamefont {Wang}}, \bibinfo {author} {\bibfnamefont {Kartik}\ \bibnamefont
  {Nayak}}, \bibinfo {author} {\bibfnamefont {Chang}\ \bibnamefont {Liu}},
  \bibinfo {author} {\bibfnamefont {T-H.~Hubert}\ \bibnamefont {Chan}},
  \bibinfo {author} {\bibfnamefont {Elaine}\ \bibnamefont {Shi}}, \bibinfo
  {author} {\bibfnamefont {Emil}\ \bibnamefont {Stefanov}}, \ and\ \bibinfo
  {author} {\bibfnamefont {Yan}\ \bibnamefont {Huang}},\ }\bibfield  {title}
  {\enquote {\bibinfo {title} {Oblivious data structures},}\ }in\ \href
  {\doibase 10.1145/2660267.2660314} {\emph {\bibinfo {booktitle} {Proceedings
  of the 2014 ACM SIGSAC Conference on Computer and Communications
  Security}}},\ \bibinfo {series and number} {CCS '14}\ (\bibinfo  {publisher}
  {ACM},\ \bibinfo {address} {New York, NY, USA},\ \bibinfo {year} {2014})\
  pp.\ \bibinfo {pages} {215--226}\BibitemShut {NoStop}%
\bibitem [{\citenamefont {Ren}\ \emph {et~al.}(2015)\citenamefont {Ren},
  \citenamefont {Fletcher}, \citenamefont {Kwon}, \citenamefont {Stefanov},
  \citenamefont {Shi}, \citenamefont {van Dijk},\ and\ \citenamefont
  {Devadas}}]{practical-oram}%
  \BibitemOpen
  \bibfield  {author} {\bibinfo {author} {\bibfnamefont {Ling}\ \bibnamefont
  {Ren}}, \bibinfo {author} {\bibfnamefont {Christopher}\ \bibnamefont
  {Fletcher}}, \bibinfo {author} {\bibfnamefont {Albert}\ \bibnamefont {Kwon}},
  \bibinfo {author} {\bibfnamefont {Emil}\ \bibnamefont {Stefanov}}, \bibinfo
  {author} {\bibfnamefont {Elaine}\ \bibnamefont {Shi}}, \bibinfo {author}
  {\bibfnamefont {Marten}\ \bibnamefont {van Dijk}}, \ and\ \bibinfo {author}
  {\bibfnamefont {Srinivas}\ \bibnamefont {Devadas}},\ }\bibfield  {title}
  {\enquote {\bibinfo {title} {Constants count: Practical improvements to
  oblivious {RAM}},}\ }in\ \href
  {https://www.usenix.org/conference/usenixsecurity15/technical-sessions/presentation/ren-ling}
  {\emph {\bibinfo {booktitle} {24th USENIX Security Symposium (USENIX Security
  15)}}}\ (\bibinfo  {publisher} {USENIX Association},\ \bibinfo {address}
  {Washington, D.C.},\ \bibinfo {year} {2015})\ pp.\ \bibinfo {pages}
  {415--430}\BibitemShut {NoStop}%
\bibitem [{\citenamefont {Ateniese}\ \emph {et~al.}(2009)\citenamefont
  {Ateniese}, \citenamefont {Benson},\ and\ \citenamefont
  {Hohenberger}}]{afgh}%
  \BibitemOpen
  \bibfield  {author} {\bibinfo {author} {\bibfnamefont {Giuseppe}\
  \bibnamefont {Ateniese}}, \bibinfo {author} {\bibfnamefont {Karyn}\
  \bibnamefont {Benson}}, \ and\ \bibinfo {author} {\bibfnamefont {Susan}\
  \bibnamefont {Hohenberger}},\ }\bibfield  {title} {\enquote {\bibinfo {title}
  {Key-private proxy re-encryption},}\ }in\ \href@noop {} {\emph {\bibinfo
  {booktitle} {Topics in Cryptology--CT-RSA 2009}}}\ (\bibinfo  {publisher}
  {Springer},\ \bibinfo {year} {2009})\ pp.\ \bibinfo {pages}
  {279--294}\BibitemShut {NoStop}%
\bibitem [{\citenamefont {Libert}\ and\ \citenamefont
  {Vergnaud}(2011)}]{libert2011unidirectional}%
  \BibitemOpen
  \bibfield  {author} {\bibinfo {author} {\bibfnamefont {Benoit}\ \bibnamefont
  {Libert}}\ and\ \bibinfo {author} {\bibfnamefont {Damien}\ \bibnamefont
  {Vergnaud}},\ }\bibfield  {title} {\enquote {\bibinfo {title} {Unidirectional
  chosen-ciphertext secure proxy re-encryption},}\ }\href@noop {} {\bibfield
  {journal} {\bibinfo  {journal} {Information Theory, IEEE Transactions on}\
  }\textbf {\bibinfo {volume} {57}},\ \bibinfo {pages} {1786--1802} (\bibinfo
  {year} {2011})}\BibitemShut {NoStop}%
\bibitem [{\citenamefont {Popa}\ and\ \citenamefont
  {Zeldovich}(2013)}]{delta-keys}%
  \BibitemOpen
  \bibfield  {author} {\bibinfo {author} {\bibfnamefont {Raluca~Ada}\
  \bibnamefont {Popa}}\ and\ \bibinfo {author} {\bibfnamefont {Nickolai}\
  \bibnamefont {Zeldovich}},\ }\href
  {https://people.csail.mit.edu/nickolai/papers/popa-multikey-eprint.pdf}
  {\enquote {\bibinfo {title} {Multi-key searchable encryption},}\ }\bibinfo
  {howpublished} {Cryptology ePrint Archive, Report 2013/508} (\bibinfo {year}
  {2013})\BibitemShut {NoStop}%
\bibitem [{\citenamefont {Popa}\ \emph {et~al.}(2014)\citenamefont {Popa},
  \citenamefont {Stark}, \citenamefont {Helfer}, \citenamefont {Valdez},
  \citenamefont {Zeldovich}, \citenamefont {Kaashoek},\ and\ \citenamefont
  {Balakrishnan}}]{mylar}%
  \BibitemOpen
  \bibfield  {author} {\bibinfo {author} {\bibfnamefont {Raluca~Ada}\
  \bibnamefont {Popa}}, \bibinfo {author} {\bibfnamefont {Emily}\ \bibnamefont
  {Stark}}, \bibinfo {author} {\bibfnamefont {Jonas}\ \bibnamefont {Helfer}},
  \bibinfo {author} {\bibfnamefont {Steven}\ \bibnamefont {Valdez}}, \bibinfo
  {author} {\bibfnamefont {Nickolai}\ \bibnamefont {Zeldovich}}, \bibinfo
  {author} {\bibfnamefont {M~Frans}\ \bibnamefont {Kaashoek}}, \ and\ \bibinfo
  {author} {\bibfnamefont {Hari}\ \bibnamefont {Balakrishnan}},\ }\bibfield
  {title} {\enquote {\bibinfo {title} {Building web applications on top of
  encrypted data using mylar},}\ }in\ \href@noop {} {\emph {\bibinfo
  {booktitle} {Proceedings of the 11th Symposium on Networked Systems Design
  and Implementation (NSDI)}}}\ (\bibinfo {year} {2014})\ pp.\ \bibinfo {pages}
  {157--172}\BibitemShut {NoStop}%
\bibitem [{\citenamefont {Weng}\ \emph {et~al.}(2009)\citenamefont {Weng},
  \citenamefont {Deng}, \citenamefont {Ding}, \citenamefont {Chu},\ and\
  \citenamefont {Lai}}]{conditional-pre-2009}%
  \BibitemOpen
  \bibfield  {author} {\bibinfo {author} {\bibfnamefont {Jian}\ \bibnamefont
  {Weng}}, \bibinfo {author} {\bibfnamefont {Robert~H.}\ \bibnamefont {Deng}},
  \bibinfo {author} {\bibfnamefont {Xuhua}\ \bibnamefont {Ding}}, \bibinfo
  {author} {\bibfnamefont {Cheng-Kang}\ \bibnamefont {Chu}}, \ and\ \bibinfo
  {author} {\bibfnamefont {Junzuo}\ \bibnamefont {Lai}},\ }\bibfield  {title}
  {\enquote {\bibinfo {title} {Conditional proxy re-encryption secure against
  chosen-ciphertext attack},}\ }in\ \href {\doibase 10.1145/1533057.1533100}
  {\emph {\bibinfo {booktitle} {Proceedings of the 4th International Symposium
  on Information, Computer, and Communications Security}}},\ \bibinfo {series
  and number} {ASIACCS '09}\ (\bibinfo  {publisher} {ACM},\ \bibinfo {address}
  {New York, NY, USA},\ \bibinfo {year} {2009})\ pp.\ \bibinfo {pages}
  {322--332}\BibitemShut {NoStop}%
\bibitem [{\citenamefont {Chow}\ \emph {et~al.}(2010)\citenamefont {Chow},
  \citenamefont {Weng}, \citenamefont {Yang},\ and\ \citenamefont
  {Deng}}]{unidirectional-pre-2010}%
  \BibitemOpen
  \bibfield  {author} {\bibinfo {author} {\bibfnamefont {ShermanS.M.}\
  \bibnamefont {Chow}}, \bibinfo {author} {\bibfnamefont {Jian}\ \bibnamefont
  {Weng}}, \bibinfo {author} {\bibfnamefont {Yanjiang}\ \bibnamefont {Yang}}, \
  and\ \bibinfo {author} {\bibfnamefont {RobertH.}\ \bibnamefont {Deng}},\
  }\bibfield  {title} {\enquote {\bibinfo {title} {Efficient unidirectional
  proxy re-encryption},}\ }in\ \href {\doibase 10.1007/978-3-642-12678-9_19}
  {\emph {\bibinfo {booktitle} {Progress in Cryptology~--- AFRICACRYPT
  2010}}},\ \bibinfo {series} {Lecture Notes in Computer Science}, Vol.\
  \bibinfo {volume} {6055},\ \bibinfo {editor} {edited by\ \bibinfo {editor}
  {\bibfnamefont {DanielJ.}\ \bibnamefont {Bernstein}}\ and\ \bibinfo {editor}
  {\bibfnamefont {Tanja}\ \bibnamefont {Lange}}}\ (\bibinfo  {publisher}
  {Springer Berlin Heidelberg},\ \bibinfo {year} {2010})\ pp.\ \bibinfo {pages}
  {316--332}\BibitemShut {NoStop}%
\bibitem [{\citenamefont {Qiu}\ \emph {et~al.}(2014)\citenamefont {Qiu},
  \citenamefont {Hwang},\ and\ \citenamefont {Lee}}]{conditional-pre-2014}%
  \BibitemOpen
  \bibfield  {author} {\bibinfo {author} {\bibfnamefont {J.}~\bibnamefont
  {Qiu}}, \bibinfo {author} {\bibfnamefont {Gi-Hyun}\ \bibnamefont {Hwang}}, \
  and\ \bibinfo {author} {\bibfnamefont {HoonJae}\ \bibnamefont {Lee}},\
  }\bibfield  {title} {\enquote {\bibinfo {title} {Efficient conditional proxy
  re-encryption with chosen-ciphertext security},}\ }in\ \href {\doibase
  10.1109/AsiaJCIS.2014.11} {\emph {\bibinfo {booktitle} {Information Security
  (ASIA JCIS), 2014 Ninth Asia Joint Conference on}}}\ (\bibinfo {year}
  {2014})\ pp.\ \bibinfo {pages} {104--110}\BibitemShut {NoStop}%
\bibitem [{\citenamefont {Fang}\ \emph {et~al.}(2012)\citenamefont {Fang},
  \citenamefont {Susilo}, \citenamefont {Ge},\ and\ \citenamefont
  {Wang}}]{hierarchical-pre}%
  \BibitemOpen
  \bibfield  {author} {\bibinfo {author} {\bibfnamefont {Liming}\ \bibnamefont
  {Fang}}, \bibinfo {author} {\bibfnamefont {Willy}\ \bibnamefont {Susilo}},
  \bibinfo {author} {\bibfnamefont {Chunpeng}\ \bibnamefont {Ge}}, \ and\
  \bibinfo {author} {\bibfnamefont {Jiandong}\ \bibnamefont {Wang}},\
  }\bibfield  {title} {\enquote {\bibinfo {title} {Hierarchical conditional
  proxy re-encryption},}\ }\href {\doibase 10.1016/j.csi.2012.01.002}
  {\bibfield  {journal} {\bibinfo  {journal} {Comput. Stand. Interfaces}\
  }\textbf {\bibinfo {volume} {34}},\ \bibinfo {pages} {380--389} (\bibinfo
  {year} {2012})}\BibitemShut {NoStop}%
\bibitem [{\citenamefont {Naveed}\ \emph {et~al.}(2015)\citenamefont {Naveed},
  \citenamefont {Kamara},\ and\ \citenamefont {Wright}}]{cryptdb-hacked}%
  \BibitemOpen
  \bibfield  {author} {\bibinfo {author} {\bibfnamefont {Muhammad}\
  \bibnamefont {Naveed}}, \bibinfo {author} {\bibfnamefont {Seny}\ \bibnamefont
  {Kamara}}, \ and\ \bibinfo {author} {\bibfnamefont {Charles~V.}\ \bibnamefont
  {Wright}},\ }\bibfield  {title} {\enquote {\bibinfo {title} {Inference
  attacks on property-preserving encrypted databases},}\ }in\ \href {\doibase
  10.1145/2810103.2813651} {\emph {\bibinfo {booktitle} {Proceedings of the
  22Nd ACM SIGSAC Conference on Computer and Communications Security}}},\
  \bibinfo {series and number} {CCS '15}\ (\bibinfo  {publisher} {ACM},\
  \bibinfo {address} {New York, NY, USA},\ \bibinfo {year} {2015})\ pp.\
  \bibinfo {pages} {644--655}\BibitemShut {NoStop}%
\bibitem [{\citenamefont {Islam}\ \emph {et~al.}(2012)\citenamefont {Islam},
  \citenamefont {Kuzu},\ and\ \citenamefont
  {Kantarcioglu}}]{access-pattern-attack}%
  \BibitemOpen
  \bibfield  {author} {\bibinfo {author} {\bibfnamefont {Mohammad~Saiful}\
  \bibnamefont {Islam}}, \bibinfo {author} {\bibfnamefont {Mehmet}\
  \bibnamefont {Kuzu}}, \ and\ \bibinfo {author} {\bibfnamefont {Murat}\
  \bibnamefont {Kantarcioglu}},\ }\bibfield  {title} {\enquote {\bibinfo
  {title} {Access pattern disclosure on searchable encryption: Ramification,
  attack and mitigation},}\ }in\ \href
  {https://www.internetsociety.org/sites/default/files/06_1.pdf} {\emph
  {\bibinfo {booktitle} {Network and Distributed System Security Symposium
  (NDSS)}}}\ (\bibinfo {year} {2012})\BibitemShut {NoStop}%
\bibitem [{us-(2014)}]{us-census-surnames}%
  \BibitemOpen
  \href
  {http://www.census.gov/topics/population/genealogy/data/2000_surnames.html}
  {\enquote {\bibinfo {title} {Frequently occurring surnames from the census
  2000},}\ }\bibinfo {howpublished} {United States Census Bureau} (\bibinfo
  {year} {2014})\BibitemShut {NoStop}%
\bibitem [{uk-(2014)}]{uk-first-names}%
  \BibitemOpen
  \href
  {http://www.ons.gov.uk/ons/publications/re-reference-tables.html?edition=tcm%3A77-370365}
  {\enquote {\bibinfo {title} {Baby names, england and wales},}\ }\bibinfo
  {howpublished} {Office for National Statistics} (\bibinfo {year}
  {2014})\BibitemShut {NoStop}%
\bibitem [{sto(2014)}]{storj}%
  \BibitemOpen
  \href {http://storj.io/storj.pdf} {\enquote {\bibinfo {title} {Storj. a
  peer-to-peer cloud storage network},}\ } (\bibinfo {year} {2014})\BibitemShut
  {NoStop}%
\bibitem [{ipf()}]{ipfs}%
  \BibitemOpen
  \href {https://ipfs.io} {\enquote {\bibinfo {title} {Ipfs is a new
  peer-to-peer hypermedia protocol},}\ }\BibitemShut {NoStop}%
\bibitem [{\citenamefont {Zyskind}\ \emph {et~al.}(2015)\citenamefont
  {Zyskind}, \citenamefont {Nathan},\ and\ \citenamefont {Pentland}}]{enigma}%
  \BibitemOpen
  \bibfield  {author} {\bibinfo {author} {\bibfnamefont {Guy}\ \bibnamefont
  {Zyskind}}, \bibinfo {author} {\bibfnamefont {Oz}~\bibnamefont {Nathan}}, \
  and\ \bibinfo {author} {\bibfnamefont {Alex}\ \bibnamefont {Pentland}},\
  }\href {http://enigma.media.mit.edu/enigma_full.pdf} {\emph {\bibinfo {title}
  {Enigma: Decentralized Computation Platform with Guaranteed Privacy}}},\
  \bibinfo {type} {Tech. Rep.}\ (\bibinfo {year} {2015})\BibitemShut {NoStop}%
\bibitem [{\citenamefont {Wikipedia}(2016{\natexlab{b}})}]{wiki:b-tree}%
  \BibitemOpen
  \bibfield  {author} {\bibinfo {author} {\bibnamefont {Wikipedia}},\ }\href
  {https://en.wikipedia.org/wiki/B-tree#Best_case_and_worst_case_heights}
  {\enquote {\bibinfo {title} {B-tree},}\ } (\bibinfo {year}
  {2016}{\natexlab{b}}),\ \bibinfo {note} {[Online; accessed
  23-Jan-2016]}\BibitemShut {NoStop}%
\bibitem [{\citenamefont {Wikipedia}(2016{\natexlab{c}})}]{wiki:lambert}%
  \BibitemOpen
  \bibfield  {author} {\bibinfo {author} {\bibnamefont {Wikipedia}},\ }\href
  {https://en.wikipedia.org/wiki/Lambert_W_function} {\enquote {\bibinfo
  {title} {Lambert {W} function},}\ } (\bibinfo {year} {2016}{\natexlab{c}}),\
  \bibinfo {note} {[Online; accessed 23-Jan-2016]}\BibitemShut {NoStop}%
\bibitem [{nie(2014)}]{nielsen-law}%
  \BibitemOpen
  \href {https://www.nngroup.com/articles/law-of-bandwidth/} {\enquote
  {\bibinfo {title} {Nielsen's law of internet bandwidth},}\ } (\bibinfo {year}
  {2014})\BibitemShut {NoStop}%
\end{thebibliography}%

\end{document}